\documentclass[aps,pra,twocolumn,superscriptaddress,longbibliography]{revtex4-1}

\usepackage{graphicx}
\usepackage{color}
\usepackage[mathlines]{lineno}
\usepackage{amsfonts}
\usepackage{amsmath}
\usepackage{nccmath}
		{\end{pmatrix}\end{medsize}}%
\usepackage{textcomp}
\usepackage{epstopdf}

\let\originalleft\left
\let\originalright\right

\renewcommand{\left}{\mathopen{}\mathclose\bgroup\originalleft}
\renewcommand{\right}{\aftergroup\egroup\originalright}

\newcommand{\bra}[1]{\ensuremath{\left\langle#1\right|}}
\newcommand{\ket}[1]{\ensuremath{\left|#1\right\rangle}}

\begin{document}
	
	
	\title{High precision Optical AC Magnetometry \\ using Dynamical Decoupling}
	
	
	
	\author{Manish K. Gupta}
	\email{\textcolor{magenta}{manishh.gupta@gmail.com}}
	\affiliation{Quantum Information and Computation Group,
		Harish-Chandra Research Institute, HBNI, Chhatnag Road, Jhunsi, Allahabad 211019, Uttar Pradesh, India}
	\affiliation{Hearne Institute for Theoretical Physics, Department of Physics and Astronomy, Louisiana State University, Baton Rouge, Louisiana 70803, USA}

	\author{Jonathan Kunjummen}
	\email{\textcolor{magenta}{jonkunjummen@gmail.com}}
	\affiliation{Homer L. Dodge Department of Physics and Astronomy, University of Oklahoma, Norman, OK 73019}
	
	\author{Xiaoting Wang}
	\affiliation{Institute of Fundamental and Frontier Sciences, University of Electronic Science and Technology of China, Chengdu, 610054, China}
	\affiliation{Hearne Institute for Theoretical Physics, Department of Physics and Astronomy, Louisiana State University, Baton Rouge, Louisiana 70803, USA}
	
	\author{Jonathan P. Dowling}
	\affiliation{Hearne Institute for Theoretical Physics, Department of Physics and Astronomy, Louisiana State University, Baton Rouge, Louisiana 70803, USA}
	\affiliation{CAS-Alibaba Quantum Computing Laboratory, USTC, Shanghai 201315, China}
	\affiliation{NYU-ECNU Institute of Physics at NYU Shanghai, Shanghai 200062, China}
	\affiliation{National Institute of Information and Communications Technology, 4-2-1, Nukui-Kitamachi, Koganei, Tokyo 184-8795, Japan}
	
	\date{\today}
	
\begin{abstract}
		We propose a magnetometer for the precise measurement of AC magnetic fields that uses a Terbium-doped optical fiber with half-waveplates built into it at specified distances.~Our scheme uses an open-loop quantum control technique called dynamical decoupling to reduce the noise floor and thus increase the sensitivity.~We show that dynamical decoupling is extremely effective in preserving the photon state evolution due to the external AC magnetic field from random birefringence in the fiber, even when accounting for errors in pulse timing.~Hence we achieve high sensitivity. For a given inter-waveplate distance, the minimum detectable field is inversely proportional to fiber length, as is the bandwidth of detectable frequencies.
\end{abstract}
	
	
	\maketitle
\section{Introduction}
\label{sec:Intro}
	
	Measurement of magnetic fields has been an active field of research for many years as it plays a pivotal role in diverse areas of science with a vast number of applications in the fields of geophysical measurements, medical imaging, military and scientific spacecraft missions. An ideal magnetometer must possess the following characteristics: ultra-high resolution, immunity to electromagnetic interference, ultra-low power consumption, a wide dynamic range and bandwidth, ultra-miniature size, low cost, and a wide range of operable temperatures \cite{Asaf17}. Even though an ideal magnetometer is a distant goal, researchers have developed a wide range of magnetometers that are optimal for specific applications. 
	
	The initial models for measurement of magnetic field were based on electromagnetic induction, but in the last few decades a variety of extremely sensitive optical magnetometry techniques have been developed that range from superconducting quantum interference devices (SQUID) \cite{Schurig07}, to cavity optomechanical system \cite{Dunlop2012,Warwick2014,Hong2017}, and spin squeezing in Bose-Einstein condensate \cite{Oberthaler2014}, and atoms \cite{Mitchell2012}. The most practical and widely used magnetometer is based on SQUIDs and achieves a sensitivity of $3.6~\textrm{fT}/\sqrt{\textrm{Hz}}$. Experimentally, the spin exchange relaxation-free magnetometry (SERF) outperforms SQUID-based magnetometers with a sensitivity of $160~\textrm{aT}/\sqrt{\textrm{Hz}}$ at room temperature \cite{Romalis2010}.~In the recent past, negatively charged nitrogen-vacancy (NV) centers in diamond \cite{Lukin2008,Budker2014,Wrachtrup2015,Jason2015} have been shown to achieve a sensitivity of $70~\textrm{pT}/\sqrt{\textrm{Hz}}$ that are nanoscale in size \cite{Budker2014}, work at room temperature, and are predicted theoretically to reach $57~\textrm{aT}/\sqrt{\textrm{Hz}}$ in sensitivity \cite{Jason2015}. A good account of the various types of magnetometer can be found in the book by Grosz et. al. \cite{Asaf17}.
	
	Optical magnetometry uses the magneto-optical effect that arises when light interacts with a medium in the presence of magnetic field. When the medium is an optical fiber, the plane of polarized light rotates as the light passes through the optical fiber and the angle of rotation is proportional to the magnetic field strength. Most high-precision magnetometers are not suitable for strong electromagnetic interference (EMI) environments because the presence of a strong ancillary electromagnetic signal causes errors in the sensors. Hence, a fiber-optic magnetometer is highly desirable as it combines the characteristics of immunity to EMI, high resolution, low power, a wide range of operable temperatures and potentially low cost. The fundamental limitation to the sensitivity of a fiber-optic magnetometer is time-varying birefringence in the fiber due to mechanical stresses and temperature fluctuations. This fluctuating birefringence increases the noise floor for the signal and thus decreases the signal-to-noise ratio (SNR) to reduce sensitivity.
	
	In this paper, we propose a novel method for high precision measurement of AC magnetic fields using dynamical decoupling in a Terbium-doped optical fiber.~We show how to apply the open-loop control protocol called dynamical decoupling (DD) in the optical fiber to decrease the noise floor of an AC signal and thus increase the sensitivity of the magnetometer. The technique for reduction of the noise floor is inspired by previous work on negatively charged nitrogen-vacancy centers in diamond that used DD to increase the coherence time of a single electron \cite{Wrachtrup}. While a lot of attention has been given to using DD to preserve the state of a qubit for the purpose of quantum computing, we apply it here to the preservation of the rotating polarization for the purpose of enhanced quantum sensing.
	
	We review the principles of optical magnetometry in section \ref{sec:OptMagnetometery}, and then in section \ref{sec:DD}, we introduce the open-loop control DD technique to preserve the coherence of quantum states. In section \ref{sec:operatingPrinciple}, we discuss the operational dynamics for control of the noise floor in AC magnetometry using DD. Finally, in section \ref{sec:NoiseModel} and \ref{sec:results}, we describe the noise model and analyze the numerical results.

\section{Optical magnetometery}
\label{sec:OptMagnetometery}
	Optical magnetometry uses the magneto-optical effect that arises when light interacts with a medium in the presence of magnetic field. The most prominent of the magneto-optical effects are the Faraday \cite{1846a,1846b,1855} and the Voigt \cite{1901} effect, i.e, rotation of light's polarization plane as it propagates through a medium placed in a longitudinal or transverse magnetic field, respectively. The Faraday rotation arises from the inequality of the refractive indices for right- and left-circularly polarized light; these, in turn, stem from the ground- and excited-state splitting in the medium when an external magnetic field is applied. The magnitude of optical rotation per unit magnetic field and unit length is characterized by the material-dependent Verdet constant $V_{\left( \lambda \tau \right)}$, which is both dispersive and temperature-dependent. The rotation of polarization azimuth $\phi_{F}$, which occurs when an optical beam propagates through a medium subjected to a Magnetic field $\bar{H}$ is given by%
	\begin{align*}
		\phi_{F}= \int_{L_{F}} V_{\left( \lambda \tau \right)} \bar{H} d \bar{l}_{F} ~,  
	\end{align*}~
	where $L_{F}$ is the interaction length \cite{RevModPhys.74.1153,Budker2013}. 
	
	The standard silica fiber has a very low Verdet constant of $1.1~\textrm{rad/(Tm)}$ at $1064$ nm, and hence a Faraday magnetometer based on standard silica fiber would be impractical due to long fiber lengths. For example, if the magnetic field is $0.2$~T, the silica fiber length required for a $45 \deg$ rotation is around $3.5$~m.~To overcome this limitation the fiber is doped with Terbium to increase the effective Verdet constant to $-32~\textrm{rad/(Tm)}$, which is $27$ times larger than that of a silica fiber \cite{Sun:10,Ballato:95}.
	
	Although the Terbium-doped optical fiber magnetometer has higher sensitivity because of large Verdet constant, the sensitivity of such an optical magnetometer is limited severely for long fiber lengths. This is due to the time-varying birefringence in the fiber caused by the mechanical stresses and temperature fluctuations that cause dephasing, or a spread in the observed polarization rotation, limiting measurement precision by increasing the noise floor and thus decreasing the SNR to reduce the sensitivity. 
	
	A technique for minimizing the quantum noise is called dynamical decoupling. The application of dynamical decoupling in an optical fiber to reduce quantum noise was first proposed by Wu and Lidar in \cite{PhysRevA.70.062310}, and later investigated in \cite{PhysRevA.85.022340,Bardhan2,Gupta1,PhysRevA.91.032329,PhysRevApplied.5.064013} for the protection of photonic qubits traveling in an optical fiber. Although dynamical decoupling has been used to preserve the state of a qubit, it has never been applied for reduction of noise in an AC signal in an optical fiber. 
		
	\section{Dynamical Decoupling}
	\label{sec:DD}
	The technique of DD is inspired by nuclear magnetic resonance (NMR), where tailored time-dependent perturbations are used to control system evolution \cite{Viola}. It is an open-loop control technique that decouples the system from environmental interactions, where the interaction is pure dephasing. It effectively preserves the dynamical evolution of the system while still minimizing the effects of the environment. 
	
	Dynamical decoupling minimizes the decoherence of the system by adding a time-dependent control Hamiltonian. If the time scale of the control is appropriately short compared to the characteristic time scale for entanglement of the system with its environment, the undesired evolution can (to limited order) be reversed, such that on periodic time scales the desired evolution of the qubit system is preserved.
	
	In DD, the control Hamiltonian is implemented by applying a sequence of pulses to the system, which is faster than the shortest time scale accessible to the reservoir degree of freedom, such that system-bath coupling is averaged to zero. The simplest pulse sequence that cancels system-environment interaction to first order is known as the Carr-Purcell-Meiboom-Gill (CPMG) DD pulse sequence \cite{CPMG,PhysRevLett.82.2417}. It is an equidistant two-pulse sequence that has been widely used for decoupling of the system from the environment.  
	
	\section{Operating Principle}
	\label{sec:operatingPrinciple}
	
	\begin{figure}[ht]
		\centering
		\fbox{\includegraphics[width=8.4cm]{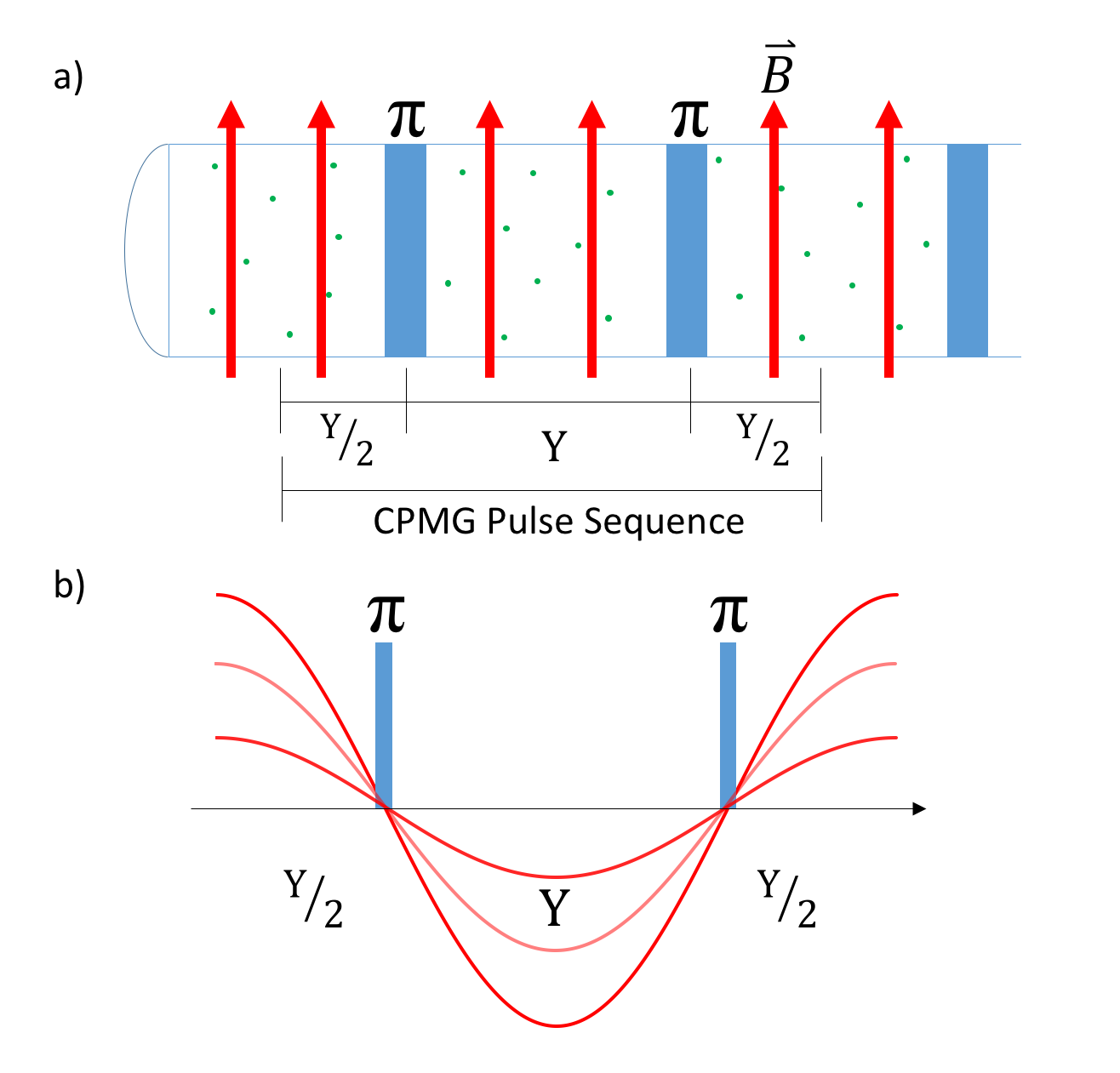}}
		\caption{(Color online) (a) Schematic of the proposed optical fiber, where the green dots represent Terbium dopants, blue bars are the half-wave plates, and red arrows show the magnetic field $\vec{B}$. The constant $Y$ denotes the distance between waveplates. The fiber implements the CPMG pulse sequence. The half-wave plate implements a $\pi$ pulse around the $x$-axis in the Bloch sphere representation of photon polarization. (b) Synchronization of AC magnetic field with one cycle of the CPMG pulse sequence such that field strength changes sign when pulse is applied.}
		\label{fig:Scheme}
	\end{figure}
	
	We propose to use a Terbium-doped optical fiber with half-wave plates built into it at specified distances for the measurement of AC magnetic fields. The schematic of the proposed scheme is shown in Fig.~\ref{fig:Scheme}(a), where the fixed inter-waveplate distance and refractive index of the optical fiber defines the characteristic measurement frequency of the device. The motivation for this sensor is its technological simplicity of design, as well as its portability and durability. The half-wave plate in the optical fiber implements the CPMG pulse sequence that helps to cancel the noise while preserving the signal i.e the rotation of photon state vector due to the external magnetic field.
	
	\begin{figure}[ht]
		\centering
		\fbox{\includegraphics[width=8.4cm]{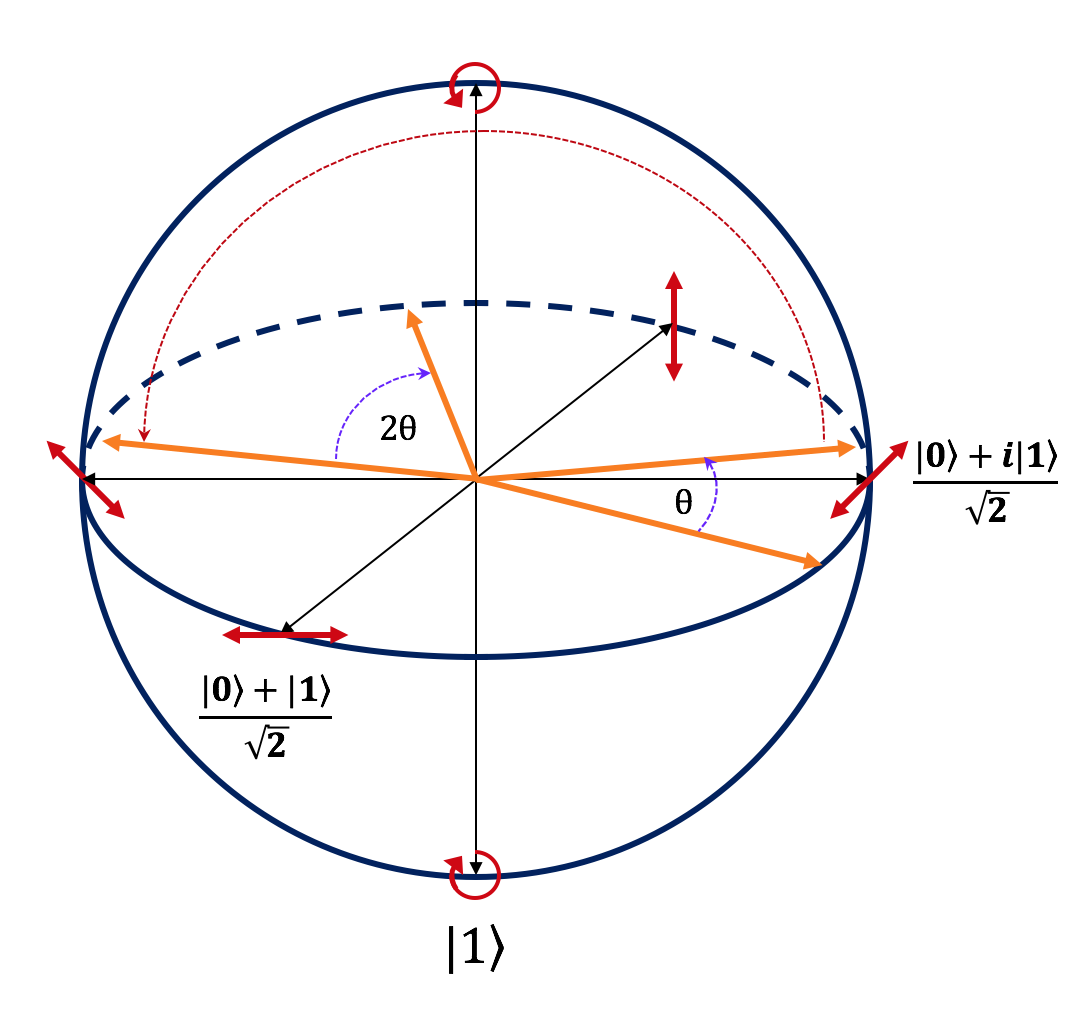}}
		\caption{(Color online) Schematic of the Bloch sphere, showing the evolution of Bloch vector in our magnetometry scheme.~In the time before application of the first pulse of the CPMG sequence, the photon polarization state rotates counterclockwise through angle $\theta$.~The pulse, a rotation by $\pi$ around the $x$-axis (dotted red arrow), is then applied.~The photon is then allowed to evolve freely through an angle $2\theta$, but since the pulse is applied at a node of the AC field, the rotation after the pulse is in the opposite direction than before the pulse.~Application of another $\pi$ pulse (not shown) after the $2\theta$ rotation will leave the state vector rotated counterclockwise by angle $3\theta$ compared to the original state. The final free evolution step (not shown) results in a state vector rotated counterclockwise by $4\theta$ from the original state.}
		\label{fig:Bloch}
	\end{figure}
	
	The quantum state of the photon propagating through the optical fiber is represented by a state vector on a Bloch Sphere as shown in Fig.~\ref{fig:Bloch}. As the photon propagates through the optical fiber placed in longitudinal AC magnetic field, the positive cycle of the magnetic field causes its state vector to rotate in the counterclockwise direction around the $z$-axis.~The negative cycle of the magnetic field would ordinarily cause the photon state vector to rotate in the opposite direction, but since the evolution of the state vector is modified by the CPMG $\pi$ pulse sequence, this instead results in the counterclockwise rotation of the state vector. The pulse sequence that was added to remove the environmental noise from the signal actually facilitates the making of the measurement itself. Since the angle of polarization rotation is proportional to the length of the fiber, all other things being equal, the sensitivity of the magnetometer is inversely proportional to the inter-waveplate distance $\tau$.
	
	Analytically, it can be shown that the maximum positive rotation occurs for photons propagating through the fiber when the magnetic field is at its maximum amplitude. The amount of polarization rotation varies sinusoidally with time at the same frequency as the field to be measured. 
	
	In general, the DD technique can not preserve the signal evolution, while eliminating the environmental noise, if the noise is in the same direction as the desired signal. Here, however, the time dependence of the field provides a means of discriminating between the signal and noise. If the control pulse is synchronized with the zero crossing of the magnetic field as shown in Fig.~\ref{fig:Scheme}(b), the angle of polarization rotation i.e. the signal can be preserved. One can view this as the field, which changes sign when the $\pi$ pulse is applied, “flipping with” the system so it is always positive in the toggling frame.
	
	\section{Noise Model}
	\label{sec:NoiseModel}
	
	Decoherence of a photonic state has its origin in fluctuating birefringence of the fiber that can result from both intrinsic and extrinsic perturbations. If a photonic qubit propagates for a length $\Delta L$ in an optical fiber then the phase accumulated by it is given by $\Delta\phi = \left(2\pi/\lambda \right)\Delta L \Delta n$, where $\Delta n$ is the birefringence of the fiber and $\lambda$ is the wavelength. We model the axially varying index dephasing in an optical fiber of length $L$ by a series of concatenated, homogeneous segments of length $\Delta L$ with constant $\Delta n$ \cite{PhysRevA.91.032329,935820,PhysRevA.85.022340}. The index fluctuations across these segments of fiber are random and the stochastic fluctuation of refractive index difference $\Delta n\left( x \right)$ across the segments are simulated as a Gaussian-distributed zero mean random process \cite{Rayleigh,935820}. The noise is defined by the first-order correlation function at two points $x_{1}$ and $x_{2}$ inside the fiber as
	
	\begin{equation}
		\left\langle \Delta n\left( x_{1} \right) \Delta n\left( x_{2} \right) \right\rangle = \exp\left[ -\Delta n\left( x \right)^{2}/2 \sigma_{\Delta n}^{2} \right].
	\end{equation}
	Here we assume that the fiber only exhibits linear index fluctuation as the radial dimension of the fiber is very small.
	
	To characterize the effectiveness of our proposed scheme, we use the fidelity $\mathfrak{F}$ between the desired output state $\ket{\psi_{d}}$, that is the final polarization state in the absence of fluctuating birefringence, and the actual output density matrix $\rho_{out}$ as
	
	\begin{equation}
		\mathfrak{F} =  \bra{\psi_{d}} \rho_{out} \ket{\psi_{d}},
	\end{equation}
	where $\rho_{out} = \frac{1}{n} \sum_{i=1}^{n} \ket{\phi_{i}}\bra{\phi_{i}}$.~Here $n$ represents the total number of randomly generated phase profiles corresponding to simulated birefringent noise in fiber and $\ket{\phi_{i}}$ is the actual photon state after traveling through the fiber.
	
	\section{Numerical Results}
	\label{sec:results}
	
	\begin{figure}[ht]
		\centering
		\fbox{\includegraphics[width=8.4cm]{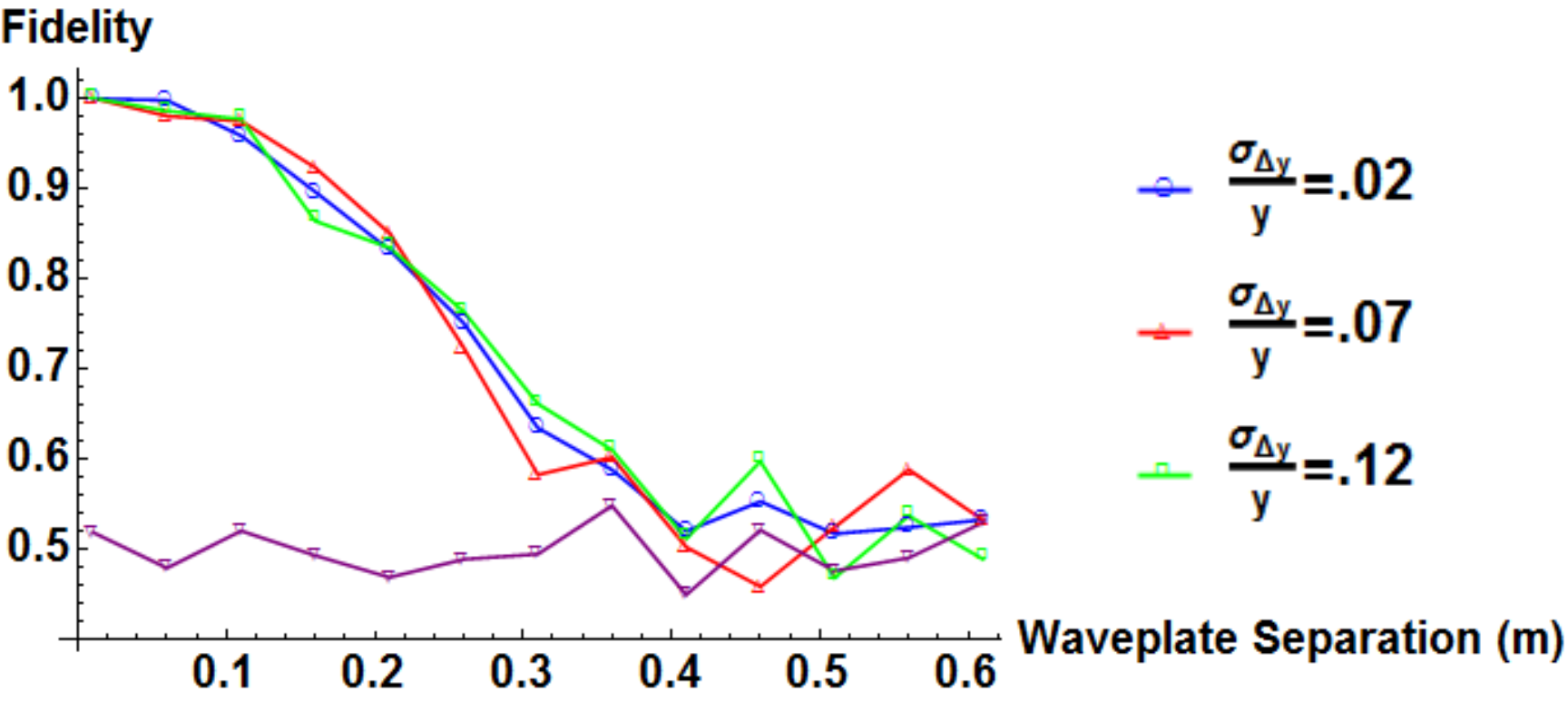}}
		\caption{(Color online) Plot of signal fidelity as a function of waveplate separation. We show fidelities without application of DD (purple curve) and with application of DD assuming three different fractional standard deviations $\sigma_{\Delta y}/y$ in wave plate placement error, where $\sigma_{\Delta y}$ is the standard deviation of error in wave plate placement and $y$ is the wave plate separation. Calculations are for a $500$~m fiber with standard error of $100$ radians in birefringence. Labeled curves use the CPMG pulse sequence.~Note that an increase in wave plate placement error does not noticeably affect fidelity for a given wave plate separation.}
		\label{fig:Fidelity vs waveplate distance} 
	\end{figure}
	We numerically demonstrate the effectiveness of our scheme in preserving the polarization rotaton of photons due to an external magnetic field. As shown in the Fig.~\ref{fig:Fidelity vs waveplate distance}, the simulation of photons traveling through the fiber in the absence of DD gives a final state fidelity of one half, indicating the photons are completely unpolarized at the end of the fiber, destroying any measurement capability. We also plot the photon fidelity, with the addition of DD pulses, and find that they do indeed, for small enough waveplate separations, preserve the desired evolution. To estimate the range of measurable magnetic field frequencies, we numerically estimate the range of inter-waveplate distance such that fidelity of the signal is greater than $95$\%. The result of the simulation is in Fig.~\ref{fig:Fidelity vs waveplate distance} that shows fidelity as a function of the inter-waveplate distance, using a fiber coherence length of $3$~m \cite{PhysRevA.85.022340}. As the inter-waveplate distance is increased, the fidelity tends toward $50$\%, which characterizes a maximally mixed final state that also means that DD is ineffective in preserving signal for larger separation of waveplates. We find that a fidelity greater than $95$\% is achieved for inter-waveplate distances less than $13$~cm. Hence taking $0.1$--$0.13$~m as the implementable range of inter-waveplate distances will put the ideal measurable magnetic field frequencies in the microwave regime, from several hundred MHz to $10$~GHz. We also see that fidelity is well preserved even for differing errors in the placement of waveplates in the optical fiber.
	
	\begin{figure}[t]
		\centering
		\fbox{\includegraphics[width=8.4cm]{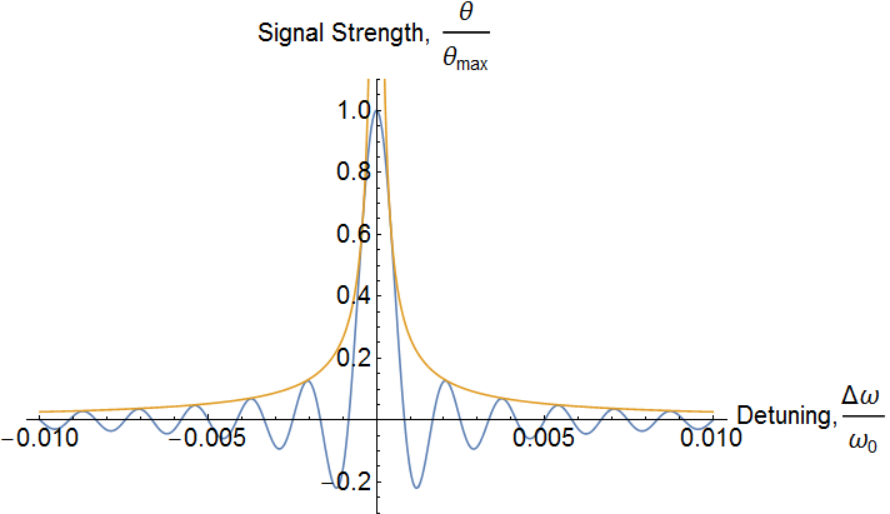}}
		\caption{(Color online) Plot shows the band of frequencies that can be measured using our design.~Because the magnetic field strength is inferred from the angle of photon polarization rotation, we define signal strength $\theta/\theta_{\textrm{max}}$ as the polarization rotation angle $\theta$ produced by a magnetic field at frequency $\omega = \Delta \omega + \omega_0$ divided by the rotation angle $\theta_{\textrm{max}}$ produced by a field at the fiber's characteristic frequency, $\omega_0$. Signal strength oscillates inside an envelope given by $\dfrac{1}{2\pi m}\dfrac{1}{\Delta \omega /\omega_0}$, where $m$ is the number of magnetic field cycles seen by the photon as it travels through the fiber.~The calculation is done for $m=600$.}
		\label{fig:Signal Strength vs Detuning}
	\end{figure} 
	
	As mentioned previously, the fixed inter-waveplate distance and the refractive index of a fiber after construction define a single frequency at which light propagating through will experience the decoupling pulses. Thus, the requirement that the pulses occur when the field is at a node implies that a given fiber will optimally measure fields oscillating at its characteristic frequency, $\omega_0=2\pi c/2yn$ where $c$ is the speed of light in vacuum, $n$ is the refractive index of the fiber, and $y$ is the inter-waveplate distance. If the field to be measured oscillates at some frequency $\omega = \Delta \omega + \omega_0$, the DD pulses will not be applied precisely at field nodes, and the error in pulse placement will accumulate as the photon travels down the fiber. Fig.~\ref{fig:Signal Strength vs Detuning} shows the dependence of signal strength on detuning from $\omega_0$ in a noiseless fiber. All loss is thus inherent in the measurement scheme. The curve has envelope $\omega_0/(2 \pi m \Delta \omega)$ where $m$ is the number of magnetic field cycles that occur in the time it takes a photon to travel through the fiber. As a result, for a given detuning, signal strength goes as one over $m$, and thus as one over fiber length.
	
	\begin{figure}[t]
		\centering
		\fbox{\includegraphics[width=8.4cm]{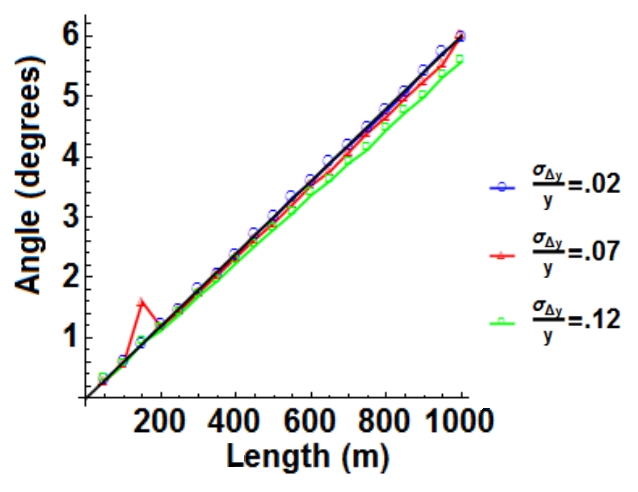}}
		\caption{(Color online) Angle of polarization rotation versus length with different waveplate placement errors. The black line is the ideal trend. The linear trend between rotation angle remains even in the presence of error in wave plate placement. Calculations use a $100$ radian birefringence profile and the CPMG pulse sequence.}
		\label{fig:linearity} 
	\end{figure}
	
	\begin{figure}[!ht]
		\centering
		\fbox{\includegraphics[width=8.4cm]{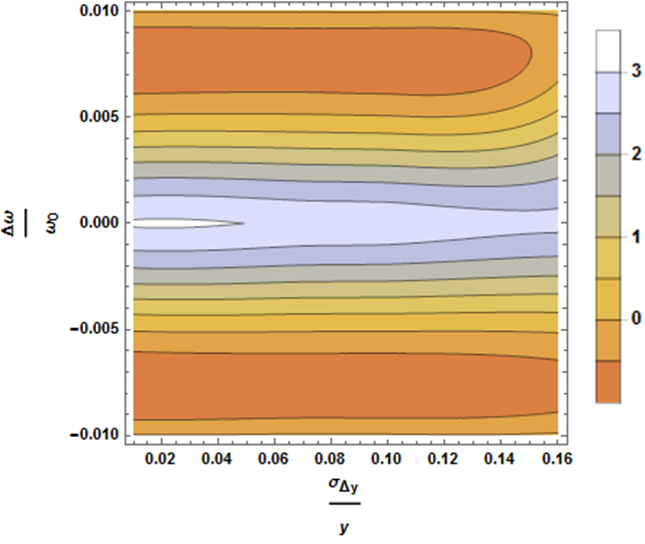}}
		\caption{(Color online) Angle of polarization rotation for a $500$~m fiber with $100$ radian birefringence profile using the CPMG pulse sequence as a function of waveplate placement error and detuning. Horizontal axis is fractional waveplate placement error and vertical axis is fractional detuning. Notice that the contours of constant polarization rotation angle remain roughly parallel as error in waveplate placement is increased. This indicates that the scheme is robust against error in waveplate placement, but also that error in waveplate placement does not result in a wider band of measurable frequencies.} 
		\label{fig:Bandwidth}      
	\end{figure}
	
	The severe dependence of the signal on synchronization also raises the question of whether the error in waveplate placement, which is to a certain extent unavoidable, undermines the viability of the scheme.~Fig.~\ref{fig:linearity} shows the angle of polarization rotation as a function of fiber length for different standard errors in waveplate placement.~The error is modeled as following a zero-mean Gaussian distribution. We see that the ideal linear relationship between the angle of rotation and path length is well preserved by dynamical decoupling, even for placement errors as high as $12$\% of the inter-waveplate distance. 
	
	In Fig.~\ref{fig:Bandwidth}, we investigate whether an increased error in waveplate placement, which results in the fiber being less tailored to its characteristic frequency, causes a  corresponding increase in the bandwidth of measurable frequencies. This, however, is not the case. The different colored bands in the figure represent regions with approximately the same angle of polarization rotation. They do not appreciably widen with increasing placement error.
	
\section{Conclusion}
\label{sec:conclusion}
We demonstrate the effectiveness of our proposed magnetometer that uses a Terbium-doped optical fiber with waveplates built into it at specified distances. The sensitivity of such a device is inversely proportional to the fiber length. This scheme remains viable even after accounting for errors in the placement of waveplates.  The sensitivity of Terbium-doped optical fiber based AC magnetometer is dependent on Verdet constant. For measuring a magnetic field of $10$~{\textmu}T with a fiber length of $4$~m, the angle of polarization rotation would be $128 \times 10^{-6}$~rad. Measurement of such small phase is possible with current metrological instruments.
	
Fiber-based magnetometry with high sensitivity has the very important military application of AC magnetic submarine field anomaly detection in the sea. Such a fiber-based magnetometer can also be used as a sensor in high voltage grids for detection of transmission failures and switchovers. Such applications give our proposed method an edge over other types of magnetometers.

	
	\begin{acknowledgments}
		The authors would like to acknowledge support from the Air Force Office of Scientific Research, the \mbox{Defense} Advanced Research Projects Agency, the Army \mbox{Research} Office, the National Science Foundation, and the Northrop Grumman Corporation.
	\end{acknowledgments}
	
	
	\bibliography{Bibliography}

\begin{thebibliography}{32}%
\makeatletter
\providecommand \@ifxundefined [1]{%
 \@ifx{#1\undefined}
}%
\providecommand \@ifnum [1]{%
 \ifnum #1\expandafter \@firstoftwo
 \else \expandafter \@secondoftwo
 \fi
}%
\providecommand \@ifx [1]{%
 \ifx #1\expandafter \@firstoftwo
 \else \expandafter \@secondoftwo
 \fi
}%
\providecommand \natexlab [1]{#1}%
\providecommand \enquote  [1]{``#1''}%
\providecommand \bibnamefont  [1]{#1}%
\providecommand \bibfnamefont [1]{#1}%
\providecommand \citenamefont [1]{#1}%
\providecommand \href@noop [0]{\@secondoftwo}%
\providecommand \href [0]{\begingroup \@sanitize@url \@href}%
\providecommand \@href[1]{\@@startlink{#1}\@@href}%
\providecommand \@@href[1]{\endgroup#1\@@endlink}%
\providecommand \@sanitize@url [0]{\catcode `\\12\catcode `\$12\catcode
  `\&12\catcode `\#12\catcode `\^12\catcode `\_12\catcode `\%12\relax}%
\providecommand \@@startlink[1]{}%
\providecommand \@@endlink[0]{}%
\providecommand \url  [0]{\begingroup\@sanitize@url \@url }%
\providecommand \@url [1]{\endgroup\@href {#1}{\urlprefix }}%
\providecommand \urlprefix  [0]{URL }%
\providecommand \Eprint [0]{\href }%
\providecommand \doibase [0]{http://dx.doi.org/}%
\providecommand \selectlanguage [0]{\@gobble}%
\providecommand \bibinfo  [0]{\@secondoftwo}%
\providecommand \bibfield  [0]{\@secondoftwo}%
\providecommand \translation [1]{[#1]}%
\providecommand \BibitemOpen [0]{}%
\providecommand \bibitemStop [0]{}%
\providecommand \bibitemNoStop [0]{.\EOS\space}%
\providecommand \EOS [0]{\spacefactor3000\relax}%
\providecommand \BibitemShut  [1]{\csname bibitem#1\endcsname}%
\let\auto@bib@innerbib\@empty
\bibitem [{\citenamefont {Grosz}\ \emph {et~al.}(2017)\citenamefont {Grosz},
  \citenamefont {Haji-Sheikh},\ and\ \citenamefont {Mukhopadhyay}}]{Asaf17}%
  \BibitemOpen
  \bibinfo {editor} {\bibfnamefont {A.}~\bibnamefont {Grosz}}, \bibinfo
  {editor} {\bibfnamefont {M.~J.}\ \bibnamefont {Haji-Sheikh}}, \ and\ \bibinfo
  {editor} {\bibfnamefont {S.~C.}\ \bibnamefont {Mukhopadhyay}},\ eds.,\ \href
  {\doibase 10.1007/978-3-319-34070-8} {\emph {\bibinfo {title} {High
  Sensitivity Magnetometers}}}\ (\bibinfo  {publisher} {Springer},\ \bibinfo
  {year} {2017})\BibitemShut {NoStop}%
\bibitem [{\citenamefont {Drung}\ \emph {et~al.}(2007)\citenamefont {Drung},
  \citenamefont {Abmann}, \citenamefont {Beyer}, \citenamefont {Kirste},
  \citenamefont {Peters}, \citenamefont {Ruede},\ and\ \citenamefont
  {Schurig}}]{Schurig07}%
  \BibitemOpen
  \bibfield  {author} {\bibinfo {author} {\bibfnamefont {D.}~\bibnamefont
  {Drung}}, \bibinfo {author} {\bibfnamefont {C.}~\bibnamefont {Abmann}},
  \bibinfo {author} {\bibfnamefont {J.}~\bibnamefont {Beyer}}, \bibinfo
  {author} {\bibfnamefont {A.}~\bibnamefont {Kirste}}, \bibinfo {author}
  {\bibfnamefont {M.}~\bibnamefont {Peters}}, \bibinfo {author} {\bibfnamefont
  {F.}~\bibnamefont {Ruede}}, \ and\ \bibinfo {author} {\bibfnamefont
  {T.}~\bibnamefont {Schurig}},\ }\href {\doibase 10.1109/TASC.2007.897403}
  {\bibfield  {journal} {\bibinfo  {journal} {IEEE Transactions on Applied
  Superconductivity}\ }\textbf {\bibinfo {volume} {17}},\ \bibinfo {pages}
  {699} (\bibinfo {year} {2007})}\BibitemShut {NoStop}%
\bibitem [{\citenamefont {Forstner}\ \emph {et~al.}(2012)\citenamefont
  {Forstner}, \citenamefont {Prams}, \citenamefont {Knittel}, \citenamefont
  {van Ooijen}, \citenamefont {Swaim}, \citenamefont {Harris}, \citenamefont
  {Szorkovszky}, \citenamefont {Bowen},\ and\ \citenamefont
  {Rubinsztein-Dunlop}}]{Dunlop2012}%
  \BibitemOpen
  \bibfield  {author} {\bibinfo {author} {\bibfnamefont {S.}~\bibnamefont
  {Forstner}}, \bibinfo {author} {\bibfnamefont {S.}~\bibnamefont {Prams}},
  \bibinfo {author} {\bibfnamefont {J.}~\bibnamefont {Knittel}}, \bibinfo
  {author} {\bibfnamefont {E.~D.}\ \bibnamefont {van Ooijen}}, \bibinfo
  {author} {\bibfnamefont {J.~D.}\ \bibnamefont {Swaim}}, \bibinfo {author}
  {\bibfnamefont {G.~I.}\ \bibnamefont {Harris}}, \bibinfo {author}
  {\bibfnamefont {A.}~\bibnamefont {Szorkovszky}}, \bibinfo {author}
  {\bibfnamefont {W.~P.}\ \bibnamefont {Bowen}}, \ and\ \bibinfo {author}
  {\bibfnamefont {H.}~\bibnamefont {Rubinsztein-Dunlop}},\ }\href {\doibase
  10.1103/PhysRevLett.108.120801} {\bibfield  {journal} {\bibinfo  {journal}
  {Phys. Rev. Lett.}\ }\textbf {\bibinfo {volume} {108}},\ \bibinfo {pages}
  {120801} (\bibinfo {year} {2012})}\BibitemShut {NoStop}%
\bibitem [{\citenamefont {Forstner}\ \emph {et~al.}(2014)\citenamefont
  {Forstner}, \citenamefont {Sheridan}, \citenamefont {Knittel}, \citenamefont
  {Humphreys}, \citenamefont {Brawley}, \citenamefont {Rubinsztein-Dunlop},\
  and\ \citenamefont {Bowen}}]{Warwick2014}%
  \BibitemOpen
  \bibfield  {author} {\bibinfo {author} {\bibfnamefont {S.}~\bibnamefont
  {Forstner}}, \bibinfo {author} {\bibfnamefont {E.}~\bibnamefont {Sheridan}},
  \bibinfo {author} {\bibfnamefont {J.}~\bibnamefont {Knittel}}, \bibinfo
  {author} {\bibfnamefont {C.~L.}\ \bibnamefont {Humphreys}}, \bibinfo {author}
  {\bibfnamefont {G.~A.}\ \bibnamefont {Brawley}}, \bibinfo {author}
  {\bibfnamefont {H.}~\bibnamefont {Rubinsztein-Dunlop}}, \ and\ \bibinfo
  {author} {\bibfnamefont {W.~P.}\ \bibnamefont {Bowen}},\ }\href {\doibase
  10.1002/adma.201401144} {\bibfield  {journal} {\bibinfo  {journal} {Advanced
  Materials}\ }\textbf {\bibinfo {volume} {26}},\ \bibinfo {pages} {6348}
  (\bibinfo {year} {2014})}\BibitemShut {NoStop}%
\bibitem [{\citenamefont {Sun}\ \emph {et~al.}(2017)\citenamefont {Sun},
  \citenamefont {Lei}, \citenamefont {Fan}, \citenamefont {Zhang},\ and\
  \citenamefont {Guo}}]{Hong2017}%
  \BibitemOpen
  \bibfield  {author} {\bibinfo {author} {\bibfnamefont {H.}~\bibnamefont
  {Sun}}, \bibinfo {author} {\bibfnamefont {Y.}~\bibnamefont {Lei}}, \bibinfo
  {author} {\bibfnamefont {S.}~\bibnamefont {Fan}}, \bibinfo {author}
  {\bibfnamefont {Q.}~\bibnamefont {Zhang}}, \ and\ \bibinfo {author}
  {\bibfnamefont {H.}~\bibnamefont {Guo}},\ }\href {\doibase
  https://doi.org/10.1016/j.physleta.2016.10.045} {\bibfield  {journal}
  {\bibinfo  {journal} {Physics Letters A}\ }\textbf {\bibinfo {volume}
  {381}},\ \bibinfo {pages} {129 } (\bibinfo {year} {2017})}\BibitemShut
  {NoStop}%
\bibitem [{\citenamefont {Muessel}\ \emph {et~al.}(2014)\citenamefont
  {Muessel}, \citenamefont {Strobel}, \citenamefont {Linnemann}, \citenamefont
  {Hume},\ and\ \citenamefont {Oberthaler}}]{Oberthaler2014}%
  \BibitemOpen
  \bibfield  {author} {\bibinfo {author} {\bibfnamefont {W.}~\bibnamefont
  {Muessel}}, \bibinfo {author} {\bibfnamefont {H.}~\bibnamefont {Strobel}},
  \bibinfo {author} {\bibfnamefont {D.}~\bibnamefont {Linnemann}}, \bibinfo
  {author} {\bibfnamefont {D.~B.}\ \bibnamefont {Hume}}, \ and\ \bibinfo
  {author} {\bibfnamefont {M.~K.}\ \bibnamefont {Oberthaler}},\ }\href
  {\doibase 10.1103/PhysRevLett.113.103004} {\bibfield  {journal} {\bibinfo
  {journal} {Phys. Rev. Lett.}\ }\textbf {\bibinfo {volume} {113}},\ \bibinfo
  {pages} {103004} (\bibinfo {year} {2014})}\BibitemShut {NoStop}%
\bibitem [{\citenamefont {Sewell}\ \emph {et~al.}(2012)\citenamefont {Sewell},
  \citenamefont {Koschorreck}, \citenamefont {Napolitano}, \citenamefont
  {Dubost}, \citenamefont {Behbood},\ and\ \citenamefont
  {Mitchell}}]{Mitchell2012}%
  \BibitemOpen
  \bibfield  {author} {\bibinfo {author} {\bibfnamefont {R.~J.}\ \bibnamefont
  {Sewell}}, \bibinfo {author} {\bibfnamefont {M.}~\bibnamefont {Koschorreck}},
  \bibinfo {author} {\bibfnamefont {M.}~\bibnamefont {Napolitano}}, \bibinfo
  {author} {\bibfnamefont {B.}~\bibnamefont {Dubost}}, \bibinfo {author}
  {\bibfnamefont {N.}~\bibnamefont {Behbood}}, \ and\ \bibinfo {author}
  {\bibfnamefont {M.~W.}\ \bibnamefont {Mitchell}},\ }\href {\doibase
  10.1103/PhysRevLett.109.253605} {\bibfield  {journal} {\bibinfo  {journal}
  {Phys. Rev. Lett.}\ }\textbf {\bibinfo {volume} {109}},\ \bibinfo {pages}
  {253605} (\bibinfo {year} {2012})}\BibitemShut {NoStop}%
\bibitem [{\citenamefont {Dang}\ \emph {et~al.}(2010)\citenamefont {Dang},
  \citenamefont {Maloof},\ and\ \citenamefont {Romalis}}]{Romalis2010}%
  \BibitemOpen
  \bibfield  {author} {\bibinfo {author} {\bibfnamefont {H.~B.}\ \bibnamefont
  {Dang}}, \bibinfo {author} {\bibfnamefont {A.~C.}\ \bibnamefont {Maloof}}, \
  and\ \bibinfo {author} {\bibfnamefont {M.~V.}\ \bibnamefont {Romalis}},\
  }\href {\doibase 10.1063/1.3491215} {\bibfield  {journal} {\bibinfo
  {journal} {Applied Physics Letters}\ }\textbf {\bibinfo {volume} {97}},\
  \bibinfo {pages} {151110} (\bibinfo {year} {2010})}\BibitemShut {NoStop}%
\bibitem [{\citenamefont {Taylor}\ \emph {et~al.}(2008)\citenamefont {Taylor},
  \citenamefont {Cappellaro}, \citenamefont {Childress}, \citenamefont {Jiang},
  \citenamefont {Budker}, \citenamefont {Hemmer}, \citenamefont {Yacoby},
  \citenamefont {Walsworth},\ and\ \citenamefont {Lukin}}]{Lukin2008}%
  \BibitemOpen
  \bibfield  {author} {\bibinfo {author} {\bibfnamefont {J.~M.}\ \bibnamefont
  {Taylor}}, \bibinfo {author} {\bibfnamefont {P.}~\bibnamefont {Cappellaro}},
  \bibinfo {author} {\bibfnamefont {L.}~\bibnamefont {Childress}}, \bibinfo
  {author} {\bibfnamefont {L.}~\bibnamefont {Jiang}}, \bibinfo {author}
  {\bibfnamefont {D.}~\bibnamefont {Budker}}, \bibinfo {author} {\bibfnamefont
  {P.~R.}\ \bibnamefont {Hemmer}}, \bibinfo {author} {\bibfnamefont
  {A.}~\bibnamefont {Yacoby}}, \bibinfo {author} {\bibfnamefont
  {R.}~\bibnamefont {Walsworth}}, \ and\ \bibinfo {author} {\bibfnamefont
  {M.~D.}\ \bibnamefont {Lukin}},\ }\href {http://dx.doi.org/10.1038/nphys1075}
  {\bibfield  {journal} {\bibinfo  {journal} {Nature Physics}\ }\textbf
  {\bibinfo {volume} {4}},\ \bibinfo {pages} {810 EP } (\bibinfo {year}
  {2008})}\BibitemShut {NoStop}%
\bibitem [{\citenamefont {Jensen}\ \emph {et~al.}(2014)\citenamefont {Jensen},
  \citenamefont {Leefer}, \citenamefont {Jarmola}, \citenamefont {Dumeige},
  \citenamefont {Acosta}, \citenamefont {Kehayias}, \citenamefont {Patton},\
  and\ \citenamefont {Budker}}]{Budker2014}%
  \BibitemOpen
  \bibfield  {author} {\bibinfo {author} {\bibfnamefont {K.}~\bibnamefont
  {Jensen}}, \bibinfo {author} {\bibfnamefont {N.}~\bibnamefont {Leefer}},
  \bibinfo {author} {\bibfnamefont {A.}~\bibnamefont {Jarmola}}, \bibinfo
  {author} {\bibfnamefont {Y.}~\bibnamefont {Dumeige}}, \bibinfo {author}
  {\bibfnamefont {V.~M.}\ \bibnamefont {Acosta}}, \bibinfo {author}
  {\bibfnamefont {P.}~\bibnamefont {Kehayias}}, \bibinfo {author}
  {\bibfnamefont {B.}~\bibnamefont {Patton}}, \ and\ \bibinfo {author}
  {\bibfnamefont {D.}~\bibnamefont {Budker}},\ }\href {\doibase
  10.1103/PhysRevLett.112.160802} {\bibfield  {journal} {\bibinfo  {journal}
  {Phys. Rev. Lett.}\ }\textbf {\bibinfo {volume} {112}},\ \bibinfo {pages}
  {160802} (\bibinfo {year} {2014})}\BibitemShut {NoStop}%
\bibitem [{\citenamefont {Wolf}\ \emph {et~al.}(2015)\citenamefont {Wolf},
  \citenamefont {Neumann}, \citenamefont {Nakamura}, \citenamefont {Sumiya},
  \citenamefont {Ohshima}, \citenamefont {Isoya},\ and\ \citenamefont
  {Wrachtrup}}]{Wrachtrup2015}%
  \BibitemOpen
  \bibfield  {author} {\bibinfo {author} {\bibfnamefont {T.}~\bibnamefont
  {Wolf}}, \bibinfo {author} {\bibfnamefont {P.}~\bibnamefont {Neumann}},
  \bibinfo {author} {\bibfnamefont {K.}~\bibnamefont {Nakamura}}, \bibinfo
  {author} {\bibfnamefont {H.}~\bibnamefont {Sumiya}}, \bibinfo {author}
  {\bibfnamefont {T.}~\bibnamefont {Ohshima}}, \bibinfo {author} {\bibfnamefont
  {J.}~\bibnamefont {Isoya}}, \ and\ \bibinfo {author} {\bibfnamefont
  {J.}~\bibnamefont {Wrachtrup}},\ }\href {\doibase 10.1103/PhysRevX.5.041001}
  {\bibfield  {journal} {\bibinfo  {journal} {Phys. Rev. X}\ }\textbf {\bibinfo
  {volume} {5}},\ \bibinfo {pages} {041001} (\bibinfo {year}
  {2015})}\BibitemShut {NoStop}%
\bibitem [{\citenamefont {Xia}\ \emph {et~al.}(2015)\citenamefont {Xia},
  \citenamefont {Zhao},\ and\ \citenamefont {Twamley}}]{Jason2015}%
  \BibitemOpen
  \bibfield  {author} {\bibinfo {author} {\bibfnamefont {K.}~\bibnamefont
  {Xia}}, \bibinfo {author} {\bibfnamefont {N.}~\bibnamefont {Zhao}}, \ and\
  \bibinfo {author} {\bibfnamefont {J.}~\bibnamefont {Twamley}} (\bibinfo
  {collaboration} {EQuS Collaboration}),\ }\href {\doibase
  10.1103/PhysRevA.92.043409} {\bibfield  {journal} {\bibinfo  {journal} {Phys.
  Rev. A}\ }\textbf {\bibinfo {volume} {92}},\ \bibinfo {pages} {043409}
  (\bibinfo {year} {2015})}\BibitemShut {NoStop}%
\bibitem [{\citenamefont {Naydenov}\ \emph {et~al.}(2011)\citenamefont
  {Naydenov}, \citenamefont {Dolde}, \citenamefont {Hall}, \citenamefont
  {Shin}, \citenamefont {Fedder}, \citenamefont {Hollenberg}, \citenamefont
  {Jelezko},\ and\ \citenamefont {Wrachtrup}}]{Wrachtrup}%
  \BibitemOpen
  \bibfield  {author} {\bibinfo {author} {\bibfnamefont {B.}~\bibnamefont
  {Naydenov}}, \bibinfo {author} {\bibfnamefont {F.}~\bibnamefont {Dolde}},
  \bibinfo {author} {\bibfnamefont {L.~T.}\ \bibnamefont {Hall}}, \bibinfo
  {author} {\bibfnamefont {C.}~\bibnamefont {Shin}}, \bibinfo {author}
  {\bibfnamefont {H.}~\bibnamefont {Fedder}}, \bibinfo {author} {\bibfnamefont
  {L.~C.~L.}\ \bibnamefont {Hollenberg}}, \bibinfo {author} {\bibfnamefont
  {F.}~\bibnamefont {Jelezko}}, \ and\ \bibinfo {author} {\bibfnamefont
  {J.}~\bibnamefont {Wrachtrup}},\ }\href {\doibase 10.1103/PhysRevB.83.081201}
  {\bibfield  {journal} {\bibinfo  {journal} {Phys. Rev. B}\ }\textbf {\bibinfo
  {volume} {83}},\ \bibinfo {pages} {081201} (\bibinfo {year}
  {2011})}\BibitemShut {NoStop}%
\bibitem [{\citenamefont {Faraday}(1846{\natexlab{a}})}]{1846a}%
  \BibitemOpen
  \bibfield  {author} {\bibinfo {author} {\bibfnamefont {M.}~\bibnamefont
  {Faraday}},\ }\href@noop {} {\bibfield  {journal} {\bibinfo  {journal}
  {Philos. Trans. R. Soc. London}\ }\textbf {\bibinfo {volume} {136}},\
  \bibinfo {pages} {1} (\bibinfo {year} {1846}{\natexlab{a}})}\BibitemShut
  {NoStop}%
\bibitem [{\citenamefont {Faraday}(1846{\natexlab{b}})}]{1846b}%
  \BibitemOpen
  \bibfield  {author} {\bibinfo {author} {\bibfnamefont {M.}~\bibnamefont
  {Faraday}},\ }\href@noop {} {\bibfield  {journal} {\bibinfo  {journal}
  {Philos. Magn.}\ }\textbf {\bibinfo {volume} {28}},\ \bibinfo {pages} {294}
  (\bibinfo {year} {1846}{\natexlab{b}})}\BibitemShut {NoStop}%
\bibitem [{\citenamefont {Faraday}(1855)}]{1855}%
  \BibitemOpen
  \bibfield  {author} {\bibinfo {author} {\bibfnamefont {M.}~\bibnamefont
  {Faraday}},\ }\href@noop {} {\bibfield  {journal} {\bibinfo  {journal}
  {Experimental Researches in Electricity}\ }\textbf {\bibinfo {volume}
  {III}},\ \bibinfo {pages} {(Taylor, London)} (\bibinfo {year}
  {1855})}\BibitemShut {NoStop}%
\bibitem [{\citenamefont {Voigt}(1901)}]{1901}%
  \BibitemOpen
  \bibfield  {author} {\bibinfo {author} {\bibfnamefont {W.}~\bibnamefont
  {Voigt}},\ }\href@noop {} {\bibfield  {journal} {\bibinfo  {journal} {Ann.
  Phys. (Leipzig)}\ }\textbf {\bibinfo {volume} {4}},\ \bibinfo {pages} {197}
  (\bibinfo {year} {1901})}\BibitemShut {NoStop}%
\bibitem [{\citenamefont {Budker}\ \emph {et~al.}(2002)\citenamefont {Budker},
  \citenamefont {Gawlik}, \citenamefont {Kimball}, \citenamefont {Rochester},
  \citenamefont {Yashchuk},\ and\ \citenamefont {Weis}}]{RevModPhys.74.1153}%
  \BibitemOpen
  \bibfield  {author} {\bibinfo {author} {\bibfnamefont {D.}~\bibnamefont
  {Budker}}, \bibinfo {author} {\bibfnamefont {W.}~\bibnamefont {Gawlik}},
  \bibinfo {author} {\bibfnamefont {D.~F.}\ \bibnamefont {Kimball}}, \bibinfo
  {author} {\bibfnamefont {S.~M.}\ \bibnamefont {Rochester}}, \bibinfo {author}
  {\bibfnamefont {V.~V.}\ \bibnamefont {Yashchuk}}, \ and\ \bibinfo {author}
  {\bibfnamefont {A.}~\bibnamefont {Weis}},\ }\href {\doibase
  10.1103/RevModPhys.74.1153} {\bibfield  {journal} {\bibinfo  {journal} {Rev.
  Mod. Phys.}\ }\textbf {\bibinfo {volume} {74}},\ \bibinfo {pages} {1153}
  (\bibinfo {year} {2002})}\BibitemShut {NoStop}%
\bibitem [{\citenamefont {Budker}\ and\ \citenamefont
  {Jackson~Kimball}(2013)}]{Budker2013}%
  \BibitemOpen
  \bibfield  {author} {\bibinfo {author} {\bibfnamefont {D.}~\bibnamefont
  {Budker}}\ and\ \bibinfo {author} {\bibfnamefont {D.~F.}\ \bibnamefont
  {Jackson~Kimball}},\ }\href@noop {} {\emph {\bibinfo {title} {Optical
  Magnetometry}}},\ \bibinfo {edition} {1st}\ ed.\ (\bibinfo  {publisher}
  {Cambridge University Press},\ \bibinfo {year} {2013})\BibitemShut {NoStop}%
\bibitem [{\citenamefont {Sun}\ \emph {et~al.}(2010)\citenamefont {Sun},
  \citenamefont {Jiang},\ and\ \citenamefont {Marciante}}]{Sun:10}%
  \BibitemOpen
  \bibfield  {author} {\bibinfo {author} {\bibfnamefont {L.}~\bibnamefont
  {Sun}}, \bibinfo {author} {\bibfnamefont {S.}~\bibnamefont {Jiang}}, \ and\
  \bibinfo {author} {\bibfnamefont {J.~R.}\ \bibnamefont {Marciante}},\ }\href
  {\doibase 10.1364/OE.18.012191} {\bibfield  {journal} {\bibinfo  {journal}
  {Opt. Express}\ }\textbf {\bibinfo {volume} {18}},\ \bibinfo {pages} {12191}
  (\bibinfo {year} {2010})}\BibitemShut {NoStop}%
\bibitem [{\citenamefont {Ballato}\ and\ \citenamefont
  {Snitzer}(1995)}]{Ballato:95}%
  \BibitemOpen
  \bibfield  {author} {\bibinfo {author} {\bibfnamefont {J.}~\bibnamefont
  {Ballato}}\ and\ \bibinfo {author} {\bibfnamefont {E.}~\bibnamefont
  {Snitzer}},\ }\href {\doibase 10.1364/AO.34.006848} {\bibfield  {journal}
  {\bibinfo  {journal} {Appl. Opt.}\ }\textbf {\bibinfo {volume} {34}},\
  \bibinfo {pages} {6848} (\bibinfo {year} {1995})}\BibitemShut {NoStop}%
\bibitem [{\citenamefont {Wu}\ and\ \citenamefont
  {Lidar}(2004)}]{PhysRevA.70.062310}%
  \BibitemOpen
  \bibfield  {author} {\bibinfo {author} {\bibfnamefont {L.-A.}\ \bibnamefont
  {Wu}}\ and\ \bibinfo {author} {\bibfnamefont {D.~A.}\ \bibnamefont {Lidar}},\
  }\href {\doibase 10.1103/PhysRevA.70.062310} {\bibfield  {journal} {\bibinfo
  {journal} {Phys. Rev. A}\ }\textbf {\bibinfo {volume} {70}},\ \bibinfo
  {pages} {062310} (\bibinfo {year} {2004})}\BibitemShut {NoStop}%
\bibitem [{\citenamefont {Roy~Bardhan}\ \emph {et~al.}(2012)\citenamefont
  {Roy~Bardhan}, \citenamefont {Anisimov}, \citenamefont {Gupta}, \citenamefont
  {Brown}, \citenamefont {Jones}, \citenamefont {Lee},\ and\ \citenamefont
  {Dowling}}]{PhysRevA.85.022340}%
  \BibitemOpen
  \bibfield  {author} {\bibinfo {author} {\bibfnamefont {B.}~\bibnamefont
  {Roy~Bardhan}}, \bibinfo {author} {\bibfnamefont {P.~M.}\ \bibnamefont
  {Anisimov}}, \bibinfo {author} {\bibfnamefont {M.~K.}\ \bibnamefont {Gupta}},
  \bibinfo {author} {\bibfnamefont {K.~L.}\ \bibnamefont {Brown}}, \bibinfo
  {author} {\bibfnamefont {N.~C.}\ \bibnamefont {Jones}}, \bibinfo {author}
  {\bibfnamefont {H.}~\bibnamefont {Lee}}, \ and\ \bibinfo {author}
  {\bibfnamefont {J.~P.}\ \bibnamefont {Dowling}},\ }\href {\doibase
  10.1103/PhysRevA.85.022340} {\bibfield  {journal} {\bibinfo  {journal} {Phys.
  Rev. A}\ }\textbf {\bibinfo {volume} {85}},\ \bibinfo {pages} {022340}
  (\bibinfo {year} {2012})}\BibitemShut {NoStop}%
\bibitem [{\citenamefont {Roy~Bardhan}\ \emph {et~al.}(2013)\citenamefont
  {Roy~Bardhan}, \citenamefont {Brown},\ and\ \citenamefont
  {Dowling}}]{Bardhan2}%
  \BibitemOpen
  \bibfield  {author} {\bibinfo {author} {\bibfnamefont {B.}~\bibnamefont
  {Roy~Bardhan}}, \bibinfo {author} {\bibfnamefont {K.~L.}\ \bibnamefont
  {Brown}}, \ and\ \bibinfo {author} {\bibfnamefont {J.~P.}\ \bibnamefont
  {Dowling}},\ }\href {\doibase 10.1103/PhysRevA.88.052311} {\bibfield
  {journal} {\bibinfo  {journal} {Phys. Rev. A}\ }\textbf {\bibinfo {volume}
  {88}},\ \bibinfo {pages} {052311} (\bibinfo {year} {2013})}\BibitemShut
  {NoStop}%
\bibitem [{\citenamefont {Gupta}\ and\ \citenamefont
  {Dowling}(2016{\natexlab{a}})}]{Gupta1}%
  \BibitemOpen
  \bibfield  {author} {\bibinfo {author} {\bibfnamefont {M.~K.}\ \bibnamefont
  {Gupta}}\ and\ \bibinfo {author} {\bibfnamefont {J.~P.}\ \bibnamefont
  {Dowling}},\ }\href {\doibase 10.1103/PhysRevApplied.5.064013} {\bibfield
  {journal} {\bibinfo  {journal} {Phys. Rev. Applied}\ }\textbf {\bibinfo
  {volume} {5}},\ \bibinfo {pages} {064013} (\bibinfo {year}
  {2016}{\natexlab{a}})}\BibitemShut {NoStop}%
\bibitem [{\citenamefont {Gupta}\ \emph {et~al.}(2015)\citenamefont {Gupta},
  \citenamefont {Navarro}, \citenamefont {Moulder}, \citenamefont {Mueller},
  \citenamefont {Balouchi}, \citenamefont {Brown}, \citenamefont {Lee},\ and\
  \citenamefont {Dowling}}]{PhysRevA.91.032329}%
  \BibitemOpen
  \bibfield  {author} {\bibinfo {author} {\bibfnamefont {M.~K.}\ \bibnamefont
  {Gupta}}, \bibinfo {author} {\bibfnamefont {E.~J.}\ \bibnamefont {Navarro}},
  \bibinfo {author} {\bibfnamefont {T.~A.}\ \bibnamefont {Moulder}}, \bibinfo
  {author} {\bibfnamefont {J.~D.}\ \bibnamefont {Mueller}}, \bibinfo {author}
  {\bibfnamefont {A.}~\bibnamefont {Balouchi}}, \bibinfo {author}
  {\bibfnamefont {K.~L.}\ \bibnamefont {Brown}}, \bibinfo {author}
  {\bibfnamefont {H.}~\bibnamefont {Lee}}, \ and\ \bibinfo {author}
  {\bibfnamefont {J.~P.}\ \bibnamefont {Dowling}},\ }\href {\doibase
  10.1103/PhysRevA.91.032329} {\bibfield  {journal} {\bibinfo  {journal} {Phys.
  Rev. A}\ }\textbf {\bibinfo {volume} {91}},\ \bibinfo {pages} {032329}
  (\bibinfo {year} {2015})}\BibitemShut {NoStop}%
\bibitem [{\citenamefont {Gupta}\ and\ \citenamefont
  {Dowling}(2016{\natexlab{b}})}]{PhysRevApplied.5.064013}%
  \BibitemOpen
  \bibfield  {author} {\bibinfo {author} {\bibfnamefont {M.~K.}\ \bibnamefont
  {Gupta}}\ and\ \bibinfo {author} {\bibfnamefont {J.~P.}\ \bibnamefont
  {Dowling}},\ }\href {\doibase 10.1103/PhysRevApplied.5.064013} {\bibfield
  {journal} {\bibinfo  {journal} {Phys. Rev. Applied}\ }\textbf {\bibinfo
  {volume} {5}},\ \bibinfo {pages} {064013} (\bibinfo {year}
  {2016}{\natexlab{b}})}\BibitemShut {NoStop}%
\bibitem [{\citenamefont {Viola}\ and\ \citenamefont {Lloyd}(1998)}]{Viola}%
  \BibitemOpen
  \bibfield  {author} {\bibinfo {author} {\bibfnamefont {L.}~\bibnamefont
  {Viola}}\ and\ \bibinfo {author} {\bibfnamefont {S.}~\bibnamefont {Lloyd}},\
  }\href {\doibase 10.1103/PhysRevA.58.2733} {\bibfield  {journal} {\bibinfo
  {journal} {Phys. Rev. A}\ }\textbf {\bibinfo {volume} {58}},\ \bibinfo
  {pages} {2733} (\bibinfo {year} {1998})}\BibitemShut {NoStop}%
\bibitem [{\citenamefont {Meiboom}\ and\ \citenamefont {Gill}(1958)}]{CPMG}%
  \BibitemOpen
  \bibfield  {author} {\bibinfo {author} {\bibfnamefont {S.}~\bibnamefont
  {Meiboom}}\ and\ \bibinfo {author} {\bibfnamefont {D.}~\bibnamefont {Gill}},\
  }\href {\doibase http://dx.doi.org/10.1063/1.1716296} {\bibfield  {journal}
  {\bibinfo  {journal} {Review of Scientific Instruments}\ }\textbf {\bibinfo
  {volume} {29}},\ \bibinfo {pages} {688} (\bibinfo {year} {1958})}\BibitemShut
  {NoStop}%
\bibitem [{\citenamefont {Viola}\ \emph {et~al.}(1999)\citenamefont {Viola},
  \citenamefont {Knill},\ and\ \citenamefont {Lloyd}}]{PhysRevLett.82.2417}%
  \BibitemOpen
  \bibfield  {author} {\bibinfo {author} {\bibfnamefont {L.}~\bibnamefont
  {Viola}}, \bibinfo {author} {\bibfnamefont {E.}~\bibnamefont {Knill}}, \ and\
  \bibinfo {author} {\bibfnamefont {S.}~\bibnamefont {Lloyd}},\ }\href
  {\doibase 10.1103/PhysRevLett.82.2417} {\bibfield  {journal} {\bibinfo
  {journal} {Phys. Rev. Lett.}\ }\textbf {\bibinfo {volume} {82}},\ \bibinfo
  {pages} {2417} (\bibinfo {year} {1999})}\BibitemShut {NoStop}%
\bibitem [{\citenamefont {Wuilpart}\ \emph {et~al.}(2001)\citenamefont
  {Wuilpart}, \citenamefont {Mégret}, \citenamefont {Blondel}, \citenamefont
  {Rogers},\ and\ \citenamefont {Defosse}}]{935820}%
  \BibitemOpen
  \bibfield  {author} {\bibinfo {author} {\bibfnamefont {M.}~\bibnamefont
  {Wuilpart}}, \bibinfo {author} {\bibfnamefont {P.}~\bibnamefont {Mégret}},
  \bibinfo {author} {\bibfnamefont {M.}~\bibnamefont {Blondel}}, \bibinfo
  {author} {\bibfnamefont {A.}~\bibnamefont {Rogers}}, \ and\ \bibinfo {author}
  {\bibfnamefont {Y.}~\bibnamefont {Defosse}},\ }\href {\doibase
  10.1109/68.935820} {\bibfield  {journal} {\bibinfo  {journal} {Photonics
  Technology Letters, IEEE}\ }\textbf {\bibinfo {volume} {13}},\ \bibinfo
  {pages} {836} (\bibinfo {year} {2001})}\BibitemShut {NoStop}%
\bibitem [{\citenamefont {Li}(1999)}]{Rayleigh}%
  \BibitemOpen
  \bibfield  {author} {\bibinfo {author} {\bibfnamefont {X.~R.}\ \bibnamefont
  {Li}},\ }\href@noop {} {\emph {\bibinfo {title} {Probability, Random Signals,
  and Statistics}}}\ (\bibinfo  {publisher} {CRC Press},\ \bibinfo {year}
  {1999})\BibitemShut {NoStop}%
\end{thebibliography}%
	
\end{document}